\documentstyle[preprint,prb,aps]{revtex}
\begin{document}
\hyphenation{smaller near studied}

\draft
\tighten
\author{V.~Bruyndoncx$^{*}$, J.~G.~Rodrigo,
T.~Puig$^{\dag}$, L.~Van~Look, V.~V.~Moshchalkov}
\address{Laboratorium voor Vaste-Stoffysica en Magnetisme, \\
Katholieke Universiteit Leuven, Celestijnenlaan 200 D,
B-3001 Leuven, Belgium}
\author{R.~Jonckheere}
\address{Interuniversity Micro-Electronics Center,
Kapeldreef 75, B-3001 Leuven, Belgium}
\title{Giant vortex state in perforated aluminum microsquares}
\date{\today }
\maketitle

\begin{abstract}
We investigate the nucleation of superconductivity
in a uniform perpendicular magnetic field H in
aluminum microsquares containing a few (2 and 4)
submicron holes (antidots). The
normal/superconducting phase boundary $T_c(H)$ of
these structures shows a quite different behavior in
low and high fields. In the low magnetic field
regime fluxoid quantization around each antidot
leads to oscillations in $T_c(H)$, expected from the
specific sample geometry, and reminiscent of the
network behavior. In high magnetic fields, the
$T_c(H)$ boundaries of the perforated and a
reference non-perforated microsquare reveal cusps at
the same values of $\Phi
/\Phi _0$ (where $\Phi $ is the applied flux
threading the total square area and $\Phi _0$ is the
superconducting flux quantum), while the background
on $T_c(H)$ becomes quasi-linear, indicating that a
giant vortex state is established. The influence of
the actual geometries on $T_c(H)$ is analyzed in the
framework of the linearized Ginzburg-Landau theory.
\end{abstract}


\draft

\section{Introduction}

Recent experiments on mesoscopic superconducting
aluminum structures with sizes smaller than the
temperature dependent coherence length $\xi(T)$ and
penetration depth $\lambda(T)$ have shown the
influence of the sample topology on the
superconducting critical parameters, like the
normal/superconducting phase boundary $T_c(H)$ (i.e.
the critical temperature $T_c$ in the presence of an
applied magnetic field $H$) \cite{Mosh1}. Many
different topologies have been studied
experimentally and theoretically.

First of all, there are structures made of
quasi-one-dimensional strips, which can be further
classified into single loops
\cite{Vloeb,daumens,bergerPV} (where $T_c(H)$ shows
the well-known periodic Little-Parks oscillations
\cite{Litt62}), multiloop structures
\cite{Vital,TP,TPPRB,dGA,Rammal} (double loop,
yin-yang, lasso, 2$\times$2-cell, etc.) and large
infinite networks \cite{Pannetier91}. The theories
used to calculate the $T_c(H)$ for these structures
are based on the linearized Ginzburg-Landau theory,
using either the de Gennes-Alexander formalism
\cite{dGA} or the London limit
\cite{Halevi83,Chi92}. The fluxoid quantization
constraint (i.e. the requirement that the order
parameter is uniquely defined after integration of
the phase gradient along a closed superconducting
contour) in these 'multiply connected' structures
gives rise to the oscillatory shape of the phase
boundaries $T_c(H)$, superimposed usually on a
parabolic background.

Secondly, surface superconductivity effects in
(circular or square shaped) single dots
\cite{Mosh1,Buisson,GeimNat,SchweigertPRL} and
antidots \cite{Bez95,BezJLTP,phyc} (i.e. one antidot
in a plain film) have been studied intensively. In
these structures, $T_c(H)$ consists of oscillations,
which are pseudoperiodic. The appearance of the
giant vortex state, where superconductivity is
nucleated only near the sample boundary
\cite{Mosh1,Buisson,SchweigertPRL,BuisSJ,GVS,qiu},
is due to the quantization of the phase winding
number $L$ of the superconducting order parameter
$\Psi=\left|\Psi\right| e^{-iL\varphi}$ (this is
equivalent to fluxoid quantization). For
cylindrically symmetric structures, one refers to
$L$ as the angular (or orbital) momentum quantum
number \cite{daumens}. For the states $L>1$, the dot
(or antidot) area is threaded by multiples $L$ of
the superconducting flux quantum $\Phi_0=h/2e$. This
"surface superconductivity" gives rise to a
quasi-linear critical field $H_{c3}$ versus
temperature $T$, which we will further compare with
the experimental $T_c(H)$.

The experimental studies of the antidot structure
are usually carried out on samples with regular
lattices of antidots. It was shown by Bezryadin {\it
et al.} that, at sufficiently high magnetic fields,
the antidots behave independently \cite{BezJLTP}. In
these systems, the antidots create efficient pinning
centers for the flux line lattice \cite{VVM96}.

We will present the measured phase boundaries
$T_c(H)$ of three different topologies, which are
shown in Fig.~\ref{atpfig1}. The three structures
studied are a filled microsquare, and two squares
with 2 and 4 square antidots respectively. Similar
structures were studied in Refs. \cite{TP,TPPRB},
where the 4-antidot structure was proposed as a
basic cell for a memory based on flux logic. In
those papers, different stable vortex configurations
were detected at low magnetic fields.

The goal of the present report is to study the
influence of the antidots inside a microsquare on
the crossover from the "network" behavior at low
fields to a giant vortex state\cite{GVS,qiu} at
higher fields, and whether eventually the two
configurations (vortices pinned by the antidots and
the giant vortex state) can coexist. We will mainly
focus on the high magnetic field regime. In a
structure with one single antidot (i.e. a loop, as
in Ref. \cite{Mosh1}) as well, the presence of a
giant vortex state can be anticipated at
sufficiently high magnetic fields. In such a system,
however, the phase winding number $L$ will be
identical for every contour encircling the antidot.
The development of the giant vortex state is
accompanied by a transition from a parabolic
background in $T_c(H)$ to a quasi-linear $T_c(H)$
behavior. The crossover field is strongly dependent
on the size and the aspect ratio of the loop and
will be the subject of a future paper. For the loop
studied in Ref. \cite{Mosh1}, the magnetic field was
clearly not sufficiently high to reach this
transition regime.

The advantage of a structure with more than one
antidot is the property that each antidot can in
principle contain a different number $L$ of flux
quanta $\Phi_0$. Simultaneously, a quantum number
$L$ is attributed to the outside square. The
observed cusps in $T_c(H)$ can then be related to
the switching of either the quantum state of an
antidot, or of the whole square.

For the 4-antidot structure, a 'collective' or network
behavior can be expected at low magnetic fields\cite{TP},
while a 'single object' regime \cite{BezJLTP} can be reached
at higher fields, where at $T_c(H)$ a surface
superconducting sheath develops near the sample boundary.
The comparison of the $T_c(H)$ data obtained on the
perforated Al microstructures with that of a reference
microsquare without antidots confirms the presence of a
giant vortex state in the three structures in the high
magnetic field regime.

\section{Experiment}

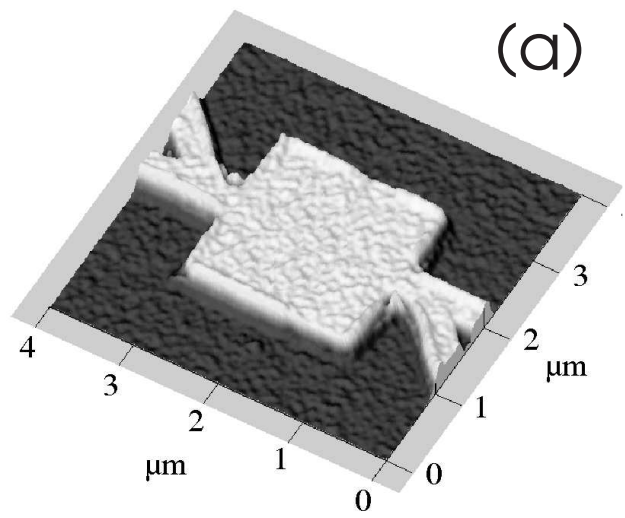
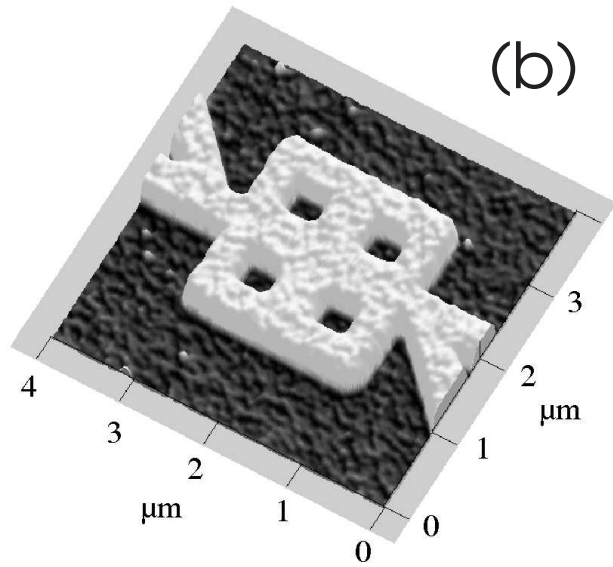
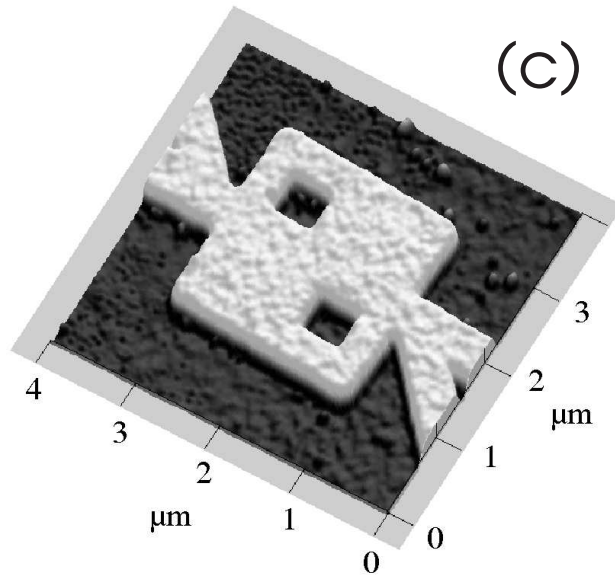
\begin{figure}[]
\caption{AFM images of the three structures:
(a) full, (b) 4-antidot, and (c) 2-antidot microsquares.}
\label{atpfig1}
\end{figure}

Three different microstructures, shown in
Fig.~\ref{atpfig1}, have been studied. A square dot,
with side $a$=2.04~$\mu m$ is taken as a reference
sample (a); a square of side $a$=2.04~$\mu m$, with
four 0.46$\times $0.46~$\mu m^2$ square antidots
(b); and a square, side $a$=2.14~$\mu m$, with two
0.52$\times $ 0.52~$\mu m^2$ antidots, placed along
a diagonal (c). For the 4-antidot sample (b) the
width of the superconducting outer stripes is
0.33~$\mu m$ and the inner stripes are 0.46~$\mu m$
wide. For the 2-antidot sample (c) the outer stripes
are 0.35 $\mu m$ wide, and the non-perforated areas
are 1.27 $\mu m$ wide. The dimensions are summarized
in Table~\ref{t1}. Electrical contacts have been
attached to the samples using an ultrasonic wire
bonding technique on the $150\times 150$~$\mu m^2$
large contact pads.

The three samples have been prepared in a single run
by thermal evaporation of $99.999\,\%$ pure Al on a
SiO$_2$ substrate. The patterns are defined using
e-beam lithography on a bilayer of PMMA resist
previous to the deposition of a $24$~nm thick
aluminum film. After the evaporation, the liftoff
was performed using dichloremethane. The structures
were characterized by X-ray, SEM and AFM
(Fig.~\ref{atpfig1}).

Four-point resistance measurements were performed in
a $^4$He cryostat, using a PAR 124A lock-in
amplifier. A measuring current of 100~nA r.m.s. with
a frequency of 27~Hz was used, which is depressing
the $T_c$ by only a few millikelvins, in the whole
magnetic field range.

The $T_c(H)$ measurements are done in a continuous run,
keeping the sample resistance typically at $50\,\%$ of the
normal state value and sweeping the magnetic field slowly
while recording the temperature. The magnetic field was
applied perpendicular to the structures, and a temperature
stability better than 0.5~mK was achieved.

\section{Results}

In Fig.~\ref{atpfig2} we present the experimental
phase boundary $T_c(H)$ of the three structures. The
measured $T_c(H)$ values were independent of the
direction of the magnetic field scans and were
reproduced in several measurement rounds. In this
paper we will always plot $T_c(H)$ in the usual way,
i.e. with the $T_c$-axis pointing from the highest
to the lowest temperature. Peaks in the $T_c(H)$
plots are then in reality local minima of the
critical temperature $T_c$.

For the reference full square, we observe
pseudoperiodic oscillations in $T_c(H)$ superimposed
with an almost linear background, where the period
of the oscillations slightly decreases with
increasing field, in agreement with previous studies
\cite{Mosh1,Buisson,GeimNat,BuisSJ,Benoist}. {\em
These observations are characteristic for the
presence of the giant vortex state.} For the
perforated microstructures, {\em two different
magnetic field regimes} can be distinguished. At
{\em high magnetic fields}, the oscillations in
$T_c(H)$ are pseudoperiodic, just as the $T_c(H)$ of
the full square. For the {\em low field} part of the
phase diagram, distinct features appear (i.e., below
$\sim 2.5$ mT): for the 2-antidot sample we observe
the same number of peaks compared to the full
square, but with a considerable shift of the
positions of the first peaks. Compared to the full
square $T_c(H)$, a new series of peaks, positioned
symmetrically with respect to $\mu_0\,H\approx 1.4$
mT, is found for the 4-antidot sample, as can be
expected for a 2$\times 2$ cell network.

In what follows, we will investigate in detail the shape of
$T_c(H)$ in the two flux regimes for the three structures.
We will discuss our results in terms of the existing models,
within the Ginzburg-Landau~(GL) theory, developed for
mesoscopic structures with a cylindrical symmetry (disks,
loops) which have been successfully applied earlier to
interpret the results obtained in mesoscopic square
structures \cite{Mosh1}.

\section{Discussion}

\subsection{The $T_c(H)$ phase boundary of the full
microsquare}

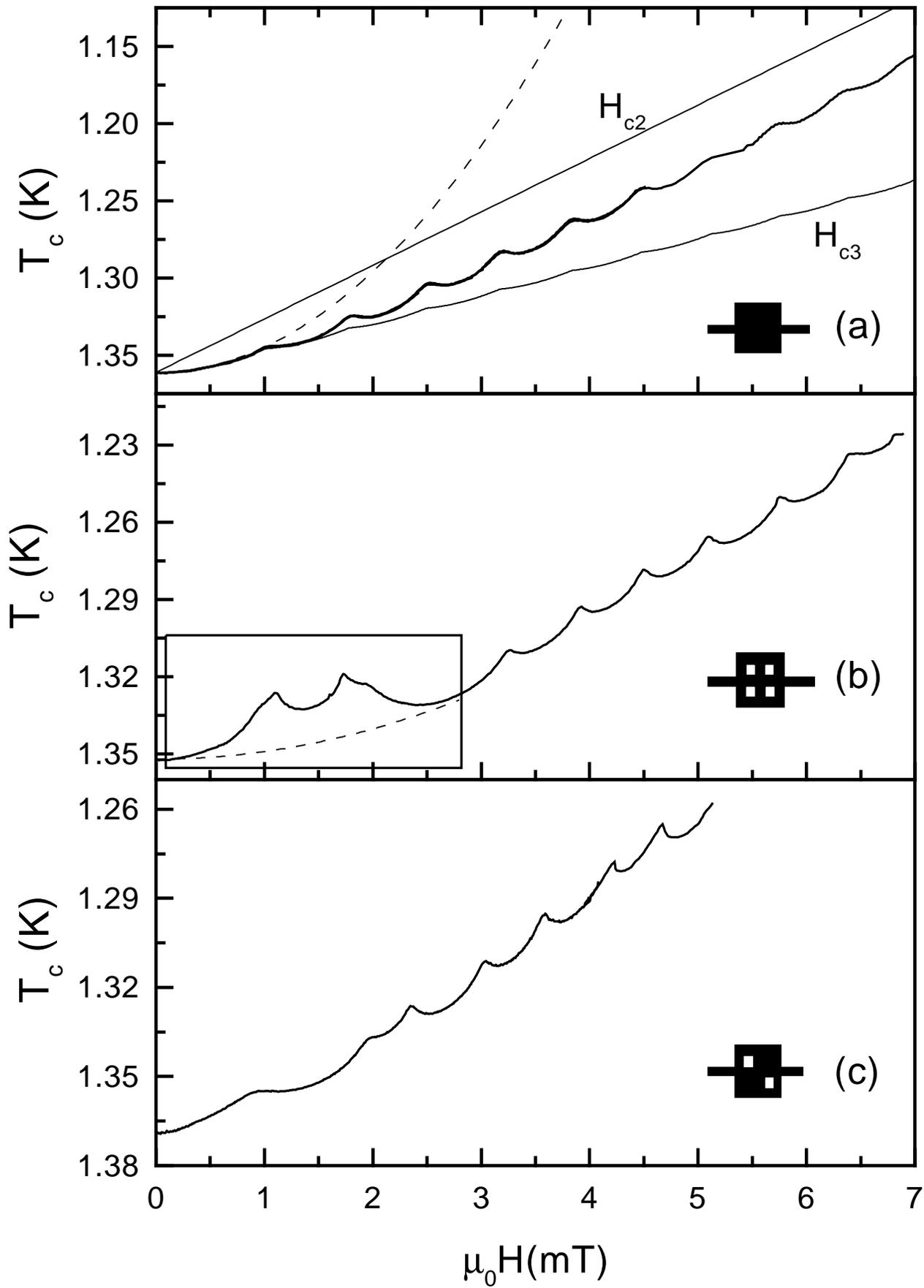
\begin{figure}[]
\caption{$T_c(H)$ phase boundaries for the (a) full,
(b) 4-antidot, and (c) 2-antidot microsquares. For
the full microsquare (a) we also present the
calculated $H_{c3}$ for an equivalent disk
(Eq.~(\ref{xitopb})). The straight solid line is the
calculated upper critical field $H_{c2}$ (substitute
$n=0$ in Eq.~(\ref{xitopb})). The parabola (dashed)
indicates the low field $T_c(H)$ behavior
(Eq.~(\ref{lowfield})). The dashed line in (b) gives
the parabolic background depression of $T_c$
(Eq.~(\ref{parabol})) for structures made of
quasi-one-dimensional strips.}
\label{atpfig2}
\end{figure}

The $T_c(H)$ curve measured for the full square
structure is very similar to the result obtained
from a calculation
\cite{SchweigertPRL,BuisSJ,qiu,Benoist} for a
mesoscopic disk in the presence of a magnetic field
(indicated as '$H_{c3}$' in Fig.~\ref{atpfig2}a). In
that model the linearized first GL~equation is
solved with the boundary condition for an ideal
superconductor/insulator interface:
\begin{equation}
\left. \left( - \imath \hbar \vec{\nabla} - 2 e
\vec{A} \right) \Psi \right|_{\perp,b}=0 \, ,
\label{Boundaryconditions}
\end{equation}
which is the condition that no supercurrent can flow
perpendicular to the interface. In the linear
approach, the vector potential $\vec{A}$ is related
to the applied magnetic field $\vec{H}$ through
$\mu_0 \, \vec{H} = rot \vec{A}$. In order to obtain
the solutions which fulfill the boundary condition
(Eq.~(\ref{Boundaryconditions})), one has to solve
the equation:
\begin{equation}
(L-\Phi/\Phi_0)\, M(-n,L+1,\Phi/\Phi_0)-\frac{2 n \,
\Phi/\Phi_0}{L+1} M(-n+1,L+2,\Phi/\Phi_0)=0
\label{Boundnum}
\end{equation}
The function $M$ is the so-called Kummer function of
the first kind, $n$ is a real number depending on
the phase winding number $L$, which has to be
obtained numerically by solving
Eq.~(\ref{Boundnum}). The flux is defined as $\Phi=
\mu_0 \, H \pi R^2$, $R$ being the radius of the disk.
The $T_c(\Phi)$ is obtained via the relation:
\begin{equation}
1-\frac{T_c(\Phi)}{T_c(0)}=4 \left( n+ \frac{1}{2} \right)
\frac{\xi^2(0)}{R^2} \frac \Phi {\Phi_0}
\label{xitopb}
\end{equation}
The upper critical field $H_{c2}$ for a bulk
superconductor is obtained when substituting $n=0$
in Eq.~(\ref{xitopb}), which gives a linear relation
between $H_{c2}$ and $T$. However, for a finite size
superconductor, a third critical field $H_{c3}$ can
be found, because the ground state is obtained from
solutions of Eq.~(\ref{Boundnum}) with $n<0$.
Superconductivity is concentrated near the sample
edge (for $L>0$), while the "normal" core contains
one or several flux quanta $L \, \Phi_0$. This
quasi-linear critical field $H_{c3}(T)$ is the
analog of the surface critical field for a
semi-infinite superconducting slab in contact with
vacuum (or insulator), where superconductivity
persists in a surface sheath up to magnetic fields
$H_{c3}(T)
\approx 1.69 \, H_{c2}(T)$ above the upper critical field
\cite{Saint-James65}.

The series of peaks in the $T_c(H)$ curve correspond
to transitions between states with different angular
momenta $L\rightarrow L+1$ of the superconducting
order parameter as successive flux quanta,
$\Phi=L\,\Phi_0$, enter the superconductor. A
comparison with the experimental result for a square
structure was made in Ref.\cite{Mosh1}. In a very
recent paper by Jadallah {\it et al.}
\cite{jadallah} the $T_c(H)$ phase boundary is
studied theoretically and is compared to the
experimental $T_c(H)$ curve for the full square,
described in the present paper.

Following Ref.\cite{Buisson}, between $\Phi $=0 and the
first peak located at $\Phi =1.92\,\,\Phi _0$, the
superconducting order parameter $\Psi$ has angular momentum
$L=0$, and the reduced critical temperature is quadratic in
$\Phi$:

\begin{equation}
1-\frac{T_c(\Phi)}{T_c(0)}=\frac{\xi^2(0)}{2R^2}\left(
\frac \Phi {\Phi _0} \right) ^2  \label{lowfield}
\end{equation}
The quasi-linear background at high flux, $\Phi
/\Phi _0\gg 1$, follows the asymptotic expression:

\begin{equation}  \label{highfield}
1-\frac{T_c(\Phi)}{T_c(0)}=\frac{2}{\eta}
 \frac{\xi^2(0)}{R^2} \frac \Phi {\Phi_0}
\end{equation}

The parameter $\eta$ represents the ratio of the ground
state energy ($H_{c3}$) to the lowest bulk Landau level
($H_{c2}$) and therefore coincides with the ratio
$H_{c3}/H_{c2}$ at a fixed temperature. For $\Phi/\Phi_0
\rightarrow \infty$, in other words for $R \rightarrow
\infty$, the value $\eta \rightarrow 1.69$. Note that
substituting $\eta=1 $ in Eq.~(\ref{highfield}) gives the
equation for $H_{c2}(T)$ (or $T_{c2}(H)$), indicated by the
straight line labeled '$H_{c2}$' in Fig.~\ref{atpfig2}a.

When matching the position of the experimental and
theoretical peaks, we obtain a value for the field
corresponding to one flux quantum, $\mu_0
\, H_0=0.53$ mT. This scaling leads to a very good
agreement in the position of all the peaks (see the
inset of Fig.~\ref{atpfig3}) and strongly supports
the validity of applying this model to our
experiments. From $H_0$ we obtain an effective area
of 3.9~$\mu m^2$, close to the actual size of the
structure, 4.2~$\mu m^2$. The introduction of this
'effective area' is obviously not needed if the
$T_c(H)$ is compared with a calculation performed
for a square \cite{jadallah}. Then, from
Eq.~(\ref{lowfield}), we find the coherence length,
$\xi (0)$=92~nm (dashed line in
Fig.~\ref{atpfig2}a), and using the values for pure
Al and the Ginzburg-Landau expressions for {\em
dirty} superconductors\cite{Tinkbook}, we can
estimate the mean free path, $\ell$=7~nm, and the
penetration depth, $\lambda (0)$=140~nm. The results
for the three structures are summarized in
Table~\ref{t1}. The determined value $\xi(0)$=92~nm
might be a bit too low, since at low magnetic fields
the electrical leads attached to the square can give
rise to nonlocal effects \cite{StrNL96}.
Simultaneously, fitting the low field part of the
experimental $T_c(H)$ to Eq.~(6) of
Ref.~\cite{jadallah} gives an increased
$\xi(0)$=95~nm.

In contrast to the experimental result presented in
Ref.\cite{Buisson}, which was obtained for a substantially
larger, but circular dot, the field period can be matched to
the theoretical predictions in the whole field interval. The
distance $\Delta \Phi $ between the peaks in $T_c(\Phi)$
follows the asymptotic limit $\Delta \Phi =\Phi _0\left(
1+\left( 2\,\eta \,\Phi /\Phi _0\right) ^{-1/2}\right) $. In
our experiment $\Delta \Phi $ reaches a nearly constant
value for $\Phi /\Phi _0\geq 6$, with $\mu
_0\,\Delta H\simeq 0.60-0.65$~mT. When a sufficiently high
magnetic field is applied to the sample, a
superconducting edge state is formed, where
superconductivity only nucleates within a surface
layer of thickness $w_H=\sqrt{\Phi _0/2\eta
\pi \mu _0\,H}$. The remaining area acts like a normal
core of radius $R_{\mbox{eff}}\approx R-w_H$, and
carries $L$ flux quanta in its interior. Due to the
expanding normal core, the sample can be seen
topologically as a loop of variable radius. For this
reason, the $T_c(H)$ of the dot shows
Little-Parks-like oscillations, which are, however,
nonperiodic. The magnetic period $\Delta H$
decreases, since the 'effective' radius grows with
increasing field. In contrast to the case of the
loop, which has a parabolic background on $T_c(H)$,
the background is quasi-linear, because of the
additional energy cost (i.e. extra reduction of
$T_c$) for depressing superconductivity in the
sample core. As the applied magnetic field grows
this 'giant vortex' core expands until it almost
fills the entire sample area. According to this
expression, at $\mu_0\,H$=5~mT, for example,
$w_H\approx 0.2$~$\mu m$ and the effective area of
the normal core for the full square structure is
$\sim $~3.3~$\mu m^2 $. This value is in agreement
with the observed magnetic period $\mu_0 \, \Delta
H$.

The amplitude of the experimental oscillations is
higher than expected from the theory (which was
observed also in Ref.\cite{Buisson}) (see
Fig.~\ref{atpfig2}a). We carried out a few $T_c(H)$
measurements, where we fixed the electronic feedback
circuit at a different resistance value ($10-90\,\%$
of the normal state resistance $R_n$). When a higher
fixed resistance value was chosen ($90\,\%$ of
$R_n$), the amplitude of the $T_c(H)$ oscillations
was decreased. We should mention here that the
measured resistance versus temperature curves, as
the magnetic field increases, reveal a 'bump' above
the normal state resistance, together with a
significant broadening of the temperature interval
in which the transition takes place. In Ref.
\cite{strunk} the bump was shown to appear in the
'superconducting state'. Since this disturbance is
still small at low fields, we believe that the
determination of $\xi(0)$ from the low field
formula~(Eq.~(\ref{lowfield})) is rather accurate
and therefore we use $\xi(0)$=92~nm further on in
this paper. The resistance peaks are caused by the
electrical leads attached to the square, which have
a higher transition temperature. For this reason,
normal/superconducting boundaries are created at $T
\approx T_c$, which, as in the experiment of Park
{\it et al.}\cite{park95}, can give rise to a
resistive transition showing a peak.

\subsection{The $T_c(H)$ phase boundaries of the
perforated microsquares}

Now, we will analyze the phase boundary observed for
the 4-antidot structure (Fig.~\ref{atpfig2}b). It
shows similarities with the full square: there is a
quasi-linear background with pseudoperiodic
oscillations at high fields and the $T_c(H)$ is
parabolic for low magnetic fields (dashed line in
Fig.~\ref{atpfig2}b). At the same time new features
(extra peaks) are clearly seen below $\sim $2.5~mT.

Let us first discuss the $T_c(H)$ curve in this low field
regime. We will compare it with the $T_c(H)$ calculated for
a $2\times 2$-cell network consisting of one-dimensional
strips. Strictly speaking, the theory for networks is valid
only when the width of the strips forming the structure are
much smaller than $\xi(T)$. For the dimensions of the sample
studied here, variations of $\Psi$ along the strip width can
be expected if $T$ is slightly below $T_c(0)$. The de
Gennes-Alexander (dGA) model\cite{dGA}, based on the
linearized Ginzburg-Landau equations, has been used
successfully to explain the phase boundaries obtained in
mesoscopic single- and multiloop
structures\cite{Mosh1,Vital} with narrow superconducting
strips. The depression of $T_c(H)$ can be expressed as the
sum of a topology dependent oscillatory component and a
parabolic term (dashed line in Fig.~\ref{atpfig2}b) which is
due to the finite width $w$ of the strips:
\begin{equation}
1-\frac{T_c(H)}{T_c(0)}=\frac{\pi ^2}3\left(
\frac{w \xi(0)\mu _0\,H} {\Phi _0}\right) ^2 \label{parabol}
\end{equation}

In Fig.~\ref{atpfig4} we compare the low magnetic
field part of the experimental $T_c(H)$, where a
parabolic background (Eq.~(\ref{parabol})) with an
averaged (over inner and outer strips) width of
0.4~$\mu m$ has been subtracted, with the $T_c(H)$
obtained from the dGA~model. In Ref.\cite{Rammal}
(Eqs.~20-22) and Ref.\cite{TPPRB} (Eq.~3.12) the
functions forming the $T_c(H)$ for a $2 \times
2$-cell network can be found. The field
corresponding to one flux quantum per elementary
cell is 2.8~mT, leading to an effective total area
for the $ 2\times 2$-cell network of 2.96~$\mu m^2$
(0.74~$\mu m^2$ per cell, the side length of each
cell $a$=0.86~$\mu m$). The theoretical $T_c(H)$
reproduce the observed flux (or fluxoid) states, and
have been calculated with $\xi(0) $=92~nm (dashed
line in Fig.~\ref{atpfig4}), obtained for the full
square, and with $\xi(0)$=140~nm (solid line in
Fig.~\ref{atpfig4}), estimated from the parabolic
background (Eq.~(\ref{parabol})).

It is clear that, for increasing magnetic field, the
$T_c(H)$ of the 4-antidot structure can no longer be
calculated from the dGA~model, since $\xi(T)$
becomes comparable to the width of the strips,
giving rise to spatial variation of $\Psi$
perpendicular to the strips.

Let us point out the differences between the
$T_c(H)$ of the present 4-antidot structure and the
previously measured '$2 \times 2$-antidot cluster'
made of Pb/Cu \cite{TP,TPPRB}. In the present Al
structure, the network features are not periodically
repeated as the magnetic field is increased.
Instead, above 2.8~mT, the positions of the
successive peaks coincide with the peaks observed in
the $T_c(H)$ of the reference full square. Moreover,
the background clearly starts deviating from
parabolic to quasi-linear. Contrary to this, in the
Pb/Cu antidot cluster (see Fig.~1 in Ref. \cite{TP})
the peaks related to the network behavior are
visible over two periods, i.e. up to $\mu_0 \, H
\approx$~5~mT. For higher fields, no giant vortex
state can be deduced for the Pb/Cu sample, since the
background reduction of $T_c$ stays parabolic and
$T_c(H)$ shows pronounced peaks instead of cusps.

The main parameter which determines the $T_c(H)$ is
the coherence length $\xi(T)$. Since the coherence
length $\xi(0)$ of Al is approximately three times
larger than for Pb/Cu, the relative $T_c$ reduction
$\delta T_c=1-T_c(H) \,/ \, T_c(0)$ is almost a
factor 10 higher in Al than in an identical Pb/Cu
sample (see Eqs.~(\ref{xitopb})-(\ref{parabol})).
Since, for a particular sample geometry, $\delta T_c
\, / \, \xi^2(0)=1 \, / \, \xi^2(T)$ should only
depend on the magnetic field (at $T=T_c(H)$), the
penetration depth $\lambda(T)$ might play an
important role. Using $\lambda^2(T)=\lambda^2(0)
\, / \,
\delta T_c$, with $\lambda(0)$=140~nm for Al, and
$\lambda(0)$=76~nm for Pb/Cu \cite{TP,TPPRB}, we
obtain $\lambda \, (T=T_{c}^{Al}) \approx 0.6 \, \,
\lambda \, (T=T_{c}^{Pb/Cu})$. In other words, the
assumption $\mu_0 \, \vec{H} = rot \vec{A}$
(corresponding to $\lambda \gg w$) is fulfilled up
to higher magnetic fields in the case of Pb/Cu. This
is consistent with the fact that, the peaks due to
switching of the state of single antidots, is seen
up to higher fields in Pb/Cu, but it does not
explain the very different behavior in the two
materials of course. Other possibilities might be
related to the proximity effect in Pb/Cu, as well as
to a different saturation number \cite{satur}
$n_s\approx R/(2\,\xi(T))$ in the two materials,
although the simple formula for $n_s$ was obtained
only for a single antidot surrounded by a large
superconducting area and might not be valid here.

For the (slightly larger) 2-antidot Al structure
(Fig.~\ref{atpfig2}c) the interpretation of the low
field regime of $T_c(H)$ is more difficult. We will
return briefly to this point later in the paper. At
high fields, however, the positions of the peaks in
$T_c(H)$ correspond to the same pseudoperiodic
oscillations as for the full square and 4-antidot
structure.

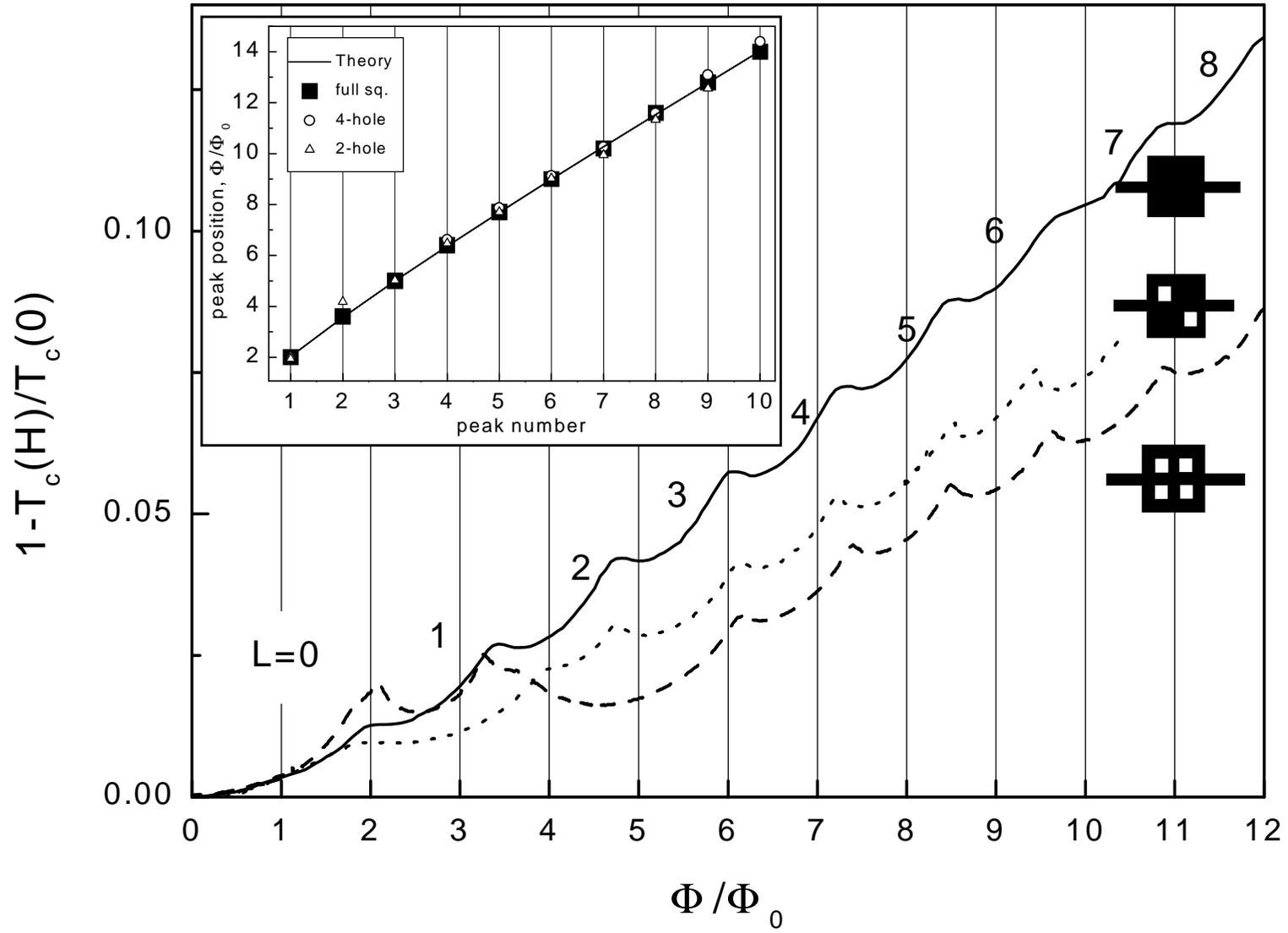
\begin{figure}[]
\caption{$T_c(\Phi)$ phase boundaries in reduced units of
critical temperature and flux. For $\Phi/\Phi_0>5$ the peaks
in $T_c(\Phi)$ appear at the same flux $\Phi/\Phi_0$ in all
the structures. The inset shows a comparison of the peak
positions in $T_c(\Phi)$ with the predictions for a
mesoscopic disk. The scaling of the theoretical values (in
flux units) gives the effective area of the microsquare.}
\label{atpfig3}
\end{figure}

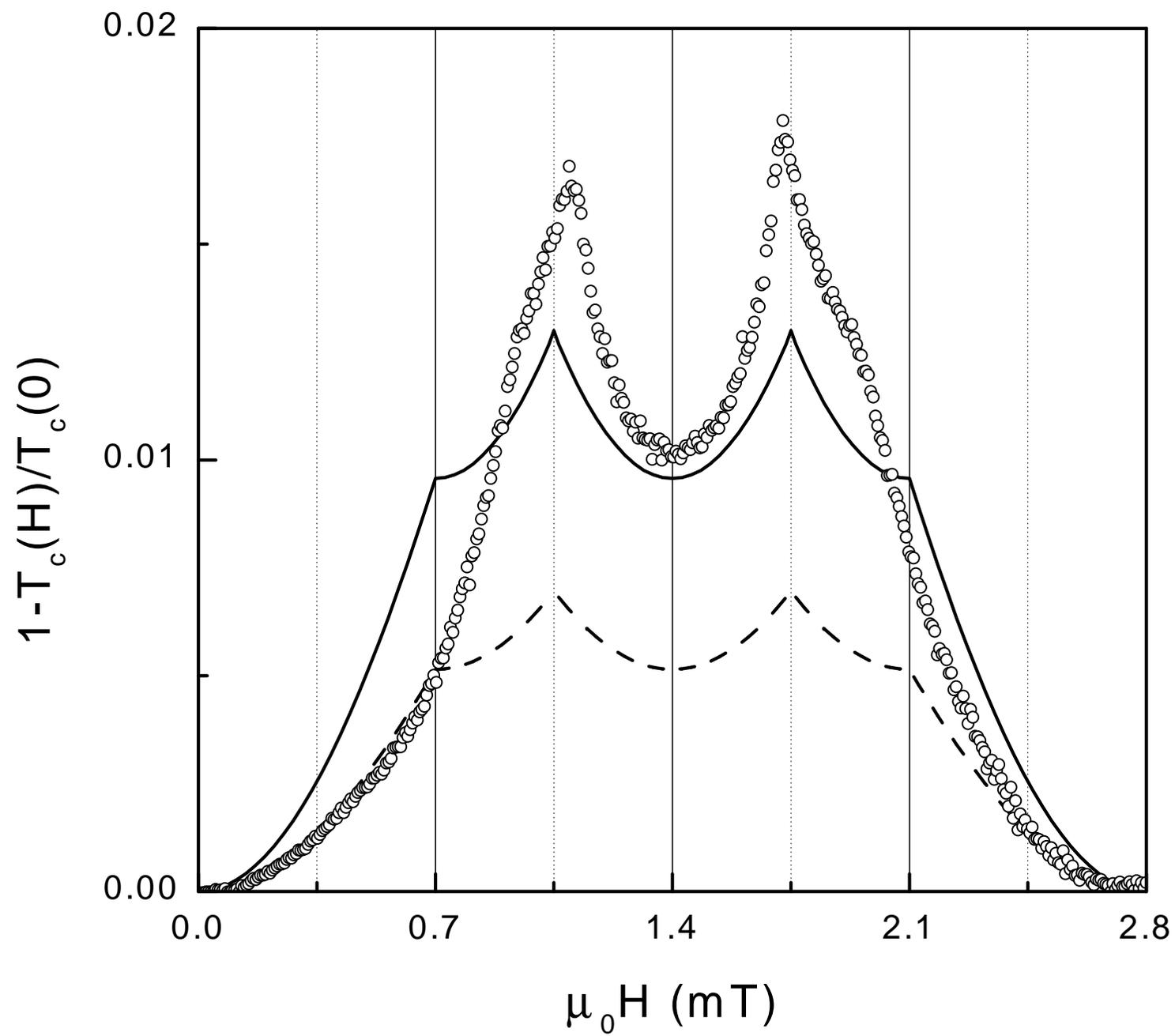
\begin{figure}[]
\caption{Low field part (single period) of the experimental
phase boundary $T_c(H)$ of the 4-antidot sample
(where a parabolic background (Eq.~(\ref{parabol}))
has been subtracted) compared with the $T_c(H)$
calculated from the dGA model for a $2 \times 2$
cell network made of one-dimensional strips (see
Refs.\protect\cite{Rammal,TPPRB}), and using
$\xi(0)$=92~nm (dashed line), obtained for the full
square, and $\xi(0)$=140~nm (solid curve), estimated
from the parabolic background (Eq.~(\ref{parabol}))
and an average strip width of 0.4 $\mu m$.}
\label{atpfig4}
\end{figure}

In Fig.~\ref{atpfig3} we have replotted the phase
diagram in units of $\Phi /\Phi_0$. It is important
to note that we defined the flux as $\Phi=\mu_0 \, H
S_{eff}$, with $S_{eff}$ the effective area of the
whole microsquare. It is close to the exact outer
sample area $S$, and was introduced in order to fit
the peak positions to the calculated $T_c(\Phi)$ for
a circular dot. To avoid confusion, we want to
mention the different definition of flux in Refs.
\cite{TP,TPPRB} (antidot cluster), where flux is
referred to the area available for one single
antidot. In the case of a loop \cite{Mosh1,StrNL96},
it is natural to define the flux as the area
enclosed by a contour through the middle of the
strips multiplied by the magnetic field, in order to
ensure a perfectly periodic $T_c(\Phi)$, with a
period $\Phi_0$.

Since the 2-antidot structure is a bit larger than
the two other structures, a different $H_0$ is used
to scale the magnetic field. In the inset the
positions of the peaks in the experimental $T_c(\Phi
)$ are compared with the theoretical prediction for
a mesoscopic superconducting disk. The $n$-th peak
corresponds to the transition between the state
$L=n-1$ and $L=n$ (for the 4-antidot sample the peak
numbers have been reassigned due to the extra peaks
in the network regime). At high magnetic fields,
there is a quite good agreement of the peak
positions of the three structures and a good
correspondence with the theoretical values found for
the disk, which is drawn in the inset of
Fig.~\ref{atpfig3} as a solid line.

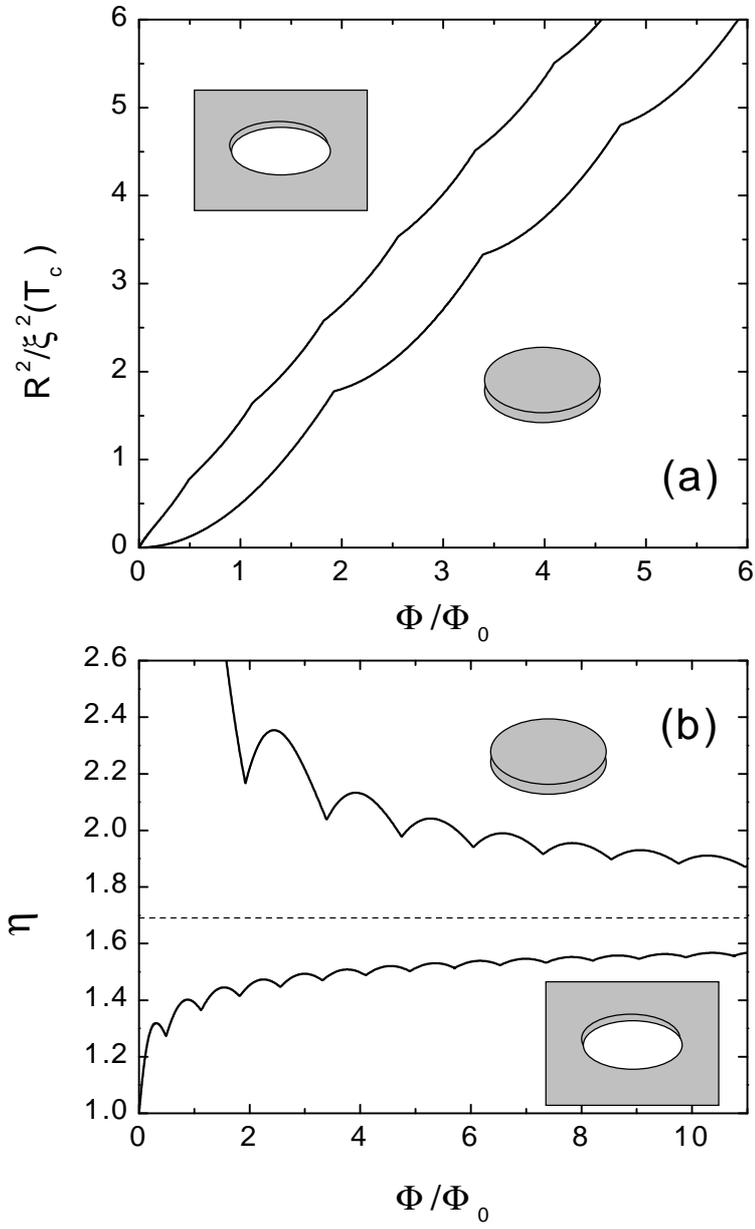
\begin{figure}[]
\caption{(a) Calculated phase boundaries (i.e.,
the $H_{c3}(T)$ curves) for a circular antidot
(Eq.~(\ref{antidotbound})) and a dot
(Eq.~(\ref{Boundnum})) in normalized units of
temperature and magnetic flux. Superconductivity
will always nucleates initially near the
dot/insulator boundary (the dot has the highest
$H_{c3}$). (b) Plot of the enhancement factor $\eta
$ (corresponding to $H_{c3}/H_{c2}$) for the same
structures.}
\label{atpfig5}
\end{figure}

How can we understand this striking coincidence of
the peak positions at high fields for the three
structures? For this, we have to look how the
superconducting order parameter nucleates along a
curved superconductor/insulator boundary.
Figure~\ref{atpfig5}a shows the calculated
$T_c(\Phi)$ curves for a single circular dot and for
an antidot (see also Ref.\cite{BezJLTP}) in an
infinite film, both of radius $R$. The latter has
been calculated in a similar way as the $T_c(\Phi)$
of the dot. For a single antidot, the boundary
condition (Eq.~(\ref{Boundaryconditions}))
translates into:
\begin{equation}
(L-\Phi/\Phi_0)\, U(-n,L+1,\Phi/\Phi_0)+ 2 n \,
\frac{\Phi}{\Phi_0} \, U(-n+1,L+2,\Phi/\Phi_0)=0 \, ,
\label{antidotbound}
\end{equation}
where the function $U$ is the Kummer function of the
second kind, diverging at the origin, i.e. at the
center of the antidot. The numeric values for $n$
have to be inserted into Eq.~(\ref{xitopb}), to
obtain the $T_c(\Phi)$.

In Figure~\ref{atpfig5}b the respective enhancement
factors $\eta $ (corresponding to $H_{c3}/H_{c2}$)
are shown. For both the dot and the antidot the
value $\eta $=1.69 is approached as the curvature
radius $R$ goes to infinity (or for a fixed $R$, as
$H \rightarrow \infty $). Since the dot has a larger
$\eta $ than the antidot, corresponding to a higher
$H_{c3}(T)$, the superconducting order parameter is
expected to grow initially at the outer sample
boundary, as the temperature drops below $T_c$. At
slightly lower temperatures surface
superconductivity should as well nucleate around the
antidots. In the mean time, however, the order
parameter has reached already a finite value over
the whole width of the strips. In the complete
temperature (or flux) interval of our measurements
$\eta <1.5$ for the antidots and $\eta>1.8$ for the
dot (when scaling the radii to the actual sample
dimensions). The resistively measured phase
transitions, probably because of this substantially
different $H_{c3}$ for a dot and an antidot, only
show peaks related to the switching of the angular
momentum $L$, associated with a closed contour along
the outer sample boundary. At the $T_c(H)$ boundary,
in the high magnetic field regime, there is no such
closed superconducting path around each single
antidot, and therefore the fluxoid quantization
condition does not need to be fulfilled for a closed
contour encircling each single antidot.

Although a more detailed analysis has to be carried
out, since the sample boundaries in our experiments
have sharp corners, the interpretation given above
is expected still to be valid. A detailed analysis
of a square {\it loop} geometry performed by Fomin
{\it et al.} \cite{Fomin} has shown that the
superconducting order parameter preferentially
nucleates near the sharp corners of the structure.
In another paper, the same authors discuss the
enhancement of the surface critical field $H_{c3}$
above the bulk upper critical field $H_{c2}$ in a
semiplane, which is bent over a certain angle
$\alpha$ (superconducting wedge) \cite{FominEPL}.
The magnetic field is parallel to the wedge edge. An
enhanced $\eta$ value is found for angles $\alpha <
\pi$ , which can be as high as $\eta=3.79$ for $\alpha=0.44
\,\pi$. For angles $\alpha > \pi$, the surface critical
field is not enhanced above $\eta=1.69$. Note that the value
$\eta=3.67$, obtained for $\alpha=\pi /2$ differs from the
calculation in Ref.~\cite{jadallah}, where for a square
domain in the limit $\Phi/\Phi_0 \rightarrow \infty$ the
factor $\eta \approx 1.8$ only. The discrepancy between
these two results still has to be clarified.

For the 2-antidot structure, further quantitative
calculations are needed to describe the low field behavior.
Since there are no extra peaks present in $T_c(\Phi)$,
compared to the full square, we believe that, here as well,
a surface superconducting sheath develops along the outer
sample boundary. At the lowest fields, the sheath width
$w_H$ is still larger than the width $w$ of the strips, and
therefore the position of the first peaks (mainly the
second) in $T_c(\Phi)$ are different from the full square.

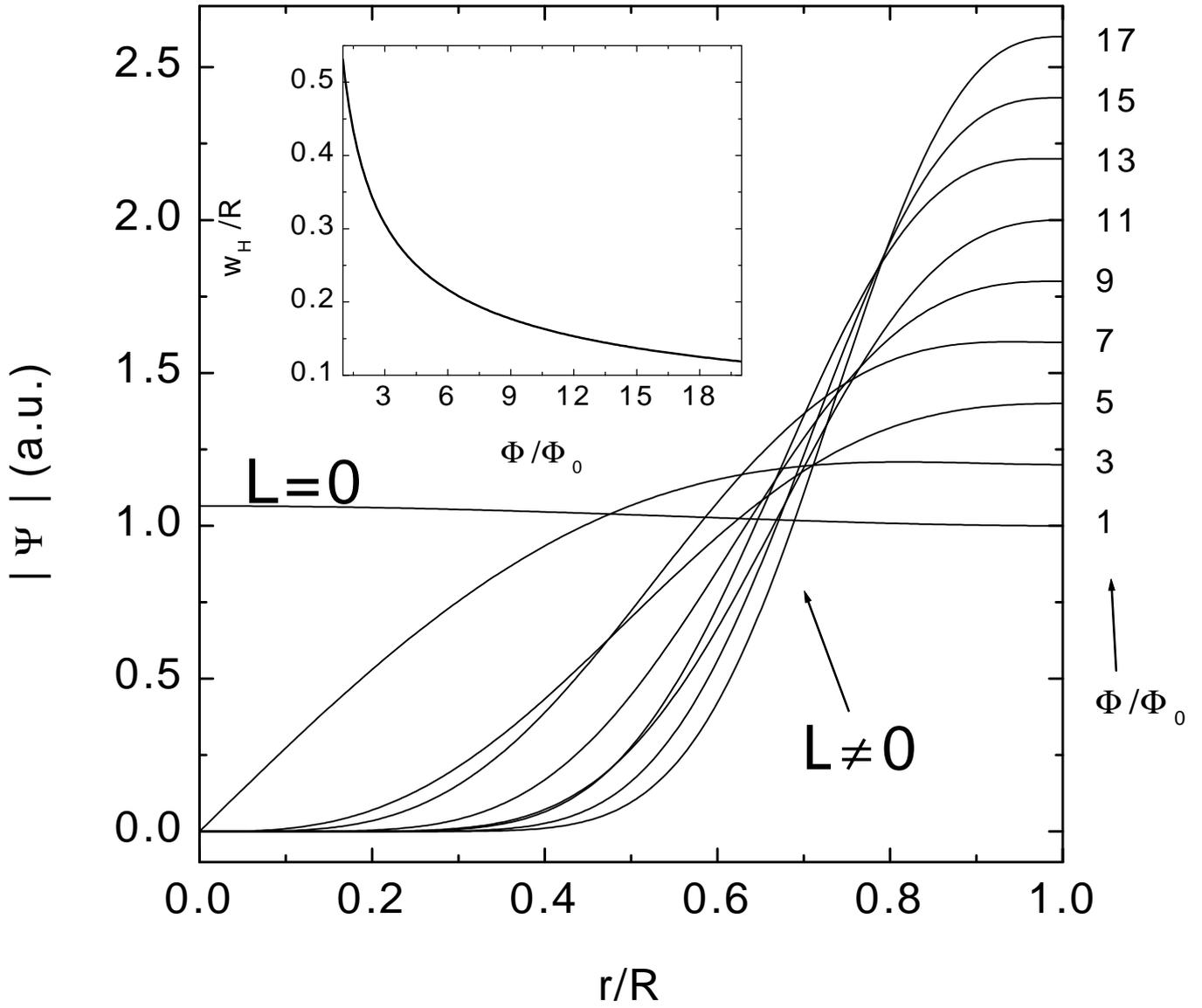
\begin{figure}[]
\caption{Calculation of the modulus of the superconducting
order parameter $|\Psi|$ for a disk, where R is the disk
radius, and r the distance measured from the sample center
to the sample boundary. The different curves are calculated
for several values for $\Phi/\Phi _0$ at the phase boundary
$H_{c3}(T)$. The inset shows the normalized width of the
superconducting edge state as a function of $\Phi/\Phi_0$.}
\label{atpfig6}
\end{figure}

In Fig.~\ref{atpfig6} the superconducting order
parameter profiles are shown for a disk, calculated
at different points $\Phi/\Phi_0=1,3,...,17$ on the
$T_c(\Phi )$ curve. For $\Phi /\Phi_0=1$ the ground
state corresponds to $L=0 $, and the order parameter
is only weakly modulated since there are no flux
lines threading the sample. As we move to higher
$\Phi /\Phi_0$ superconductivity becomes more and
more concentrated near the sample boundary. The
presence of the antidots in the perforated samples
produces different profiles for the superconducting
order parameter, since the $\Psi$ has to fulfill the
boundary conditions (Eq.~(\ref{Boundaryconditions}))
also at the antidot boundaries. A two-dimensional
GL~calculation would be required to obtain the
proper order parameter distribution here.

The inset of Fig.~\ref{atpfig6} shows the width
$w_H$ of the edge state as a function of $\Phi
/\Phi _0$. For our samples $w_H$ becomes equal to the width
of the strips at $\sim (3-4)\,\Phi /\Phi _0$. For
fluxes above this value the presence of the antidots
will not influence the position of the peaks in
$T_c(\Phi)$. Therefore, at high fields, when $w_H$
is smaller than the width of the outer strips in our
structures, the order parameter is restricted to the
outer border of the sample, and therefore it is
impossible to have supercurrents around a single
antidot.

The background depression of $T_c$ is different for
the three structures studied (Fig.~\ref{atpfig3}).
The larger the perforated area (in other words the
smaller the area exposed to the perpendicular
magnetic field), the less $T_c(\Phi )$ is pushed to
lower temperatures. Another clear example of a
similar behavior is given in Ref. \cite{Mosh1},
where the $T_c(H)$ of the (square) dot is shown to
be lower than the $T_c(H)$ of a the loop, when
exposed to a perpendicular magnetic field. This
general rule applies, for instance, also to simple
strips for which $T_c(H)$ is suppressed more when
the width $w$ increases, which is described by
Eq.~(\ref{parabol}). In the dot of Ref.
\cite{Mosh1}, the giant vortex state was shown to
develop. For the loop studied in that paper
\cite{Mosh1}, however, the used magnetic fields were
too low to induce the crossover to a giant vortex
state, which contrasts the observations from the
present paper, in the high magnetic field regime.

The appearance of the giant vortex state in the high
field regime is the most plausible explanation at
the moment. We can not exclude, however, that
another scenario is also possible. Namely, a nearly
flat non-zero distribution of $\left| \Psi \right| $
in the sample interior could coexist with an
enhanced $\left|\Psi \right| $ at the external
sample boundary, although we believe that such a
situation would give rise to peaks in $T_c(\Phi)$
each time an antidot changes its quantum state. In
any case, the final description of the specific
shape of the superconducting order parameter at
$T_c(\Phi)$ requires a numerical two-dimensional
calculation of $\Psi$ for the perforated topologies,
where the boundary conditions are fulfilled both at
the outer and at antidot superconducting/insulator
interfaces.

In summary, we have presented the experimental
superconducting/normal phase boundaries $T_c(H)$ of
a mesoscopic full square and two perforated
mesoscopic aluminum squares. The flux interval was
divided in two regimes by comparing the results with
the behavior of a full square microstructure: for
low magnetic fields the 4-antidot structure behaves
like a network consisting of quasi-one-dimensional
strips, giving rise to extra peaks in $T_c(H)$ in
comparison to the full square. In the 2-antidot
structure the peak positions are only shifted
compared to the full square. As soon as each antidot
contains one flux quantum, the giant vortex
develops, resulting in pseudoperiodic oscillations
in the $T_c(H)$ and a quasi-linear background on
$T_c(H)$ at high magnetic fields. In this regime,
the peak positions coincide for all three structures
studied when the phase boundaries are plotted in
flux quanta units (where flux is referred to the
total sample area). For high magnetic fields, the
presence of the antidots apparently does not change
the phase winding number $L$ for a closed contour
around the outer perimeter of the whole square.
Since the enhancement factor $H_{c3}/H_{c2}$ is the
highest at the outer sample boundary,
superconductivity nucleates initially near the
outside sample edges, resulting in a giant vortex
state.

{\em Note added in proof:} The discrepancy between
the result obtained by the authors of
Ref.~\cite{jadallah} and of Ref.~\cite{FominEPL} has
been clarified in an erratum~\cite{FominEPLerrat}.

\section*{Acknowledgments}

The authors are thankful to the FWO-Vlaanderen, the
Flemish Concerted Action (GOA) and the Belgian
Inter-University Attraction Poles (IUAP) for the
financial support. T.~Puig wishes to thank the
Training and Mobility of Researchers Program of the
European Union. J.~G.~Rodrigo is a Research Fellow
of the K.U.Leuven Onderzoeksraad. Discussions with
Y.~Bruynseraede, V.M.~Fomin, J.~Devreese,
J.~Rubinstein, C.~Strunk and E.~Rosseel are
gratefully acknowledged.
\\

\noindent $^{*}$ e-mail: Vital.Bruyndoncx@fys.kuleuven.ac.be \\
$^{\dag}$ Present address: Institut de Ciencia de Materials
de Barcelona - CSIC, Campus de la UAB, 08193 Bellaterra,
Spain.

\begin{table}
\begin{tabular}{|c|c|c|c|c|c|c|c|}
\hline
{\bf Low H} & area ($\mu $m$^2$) & $T_c(0)$ (K) &
$\mu_0H_0$ (mT) & eff. area ($\mu $m$^2$) &
$\xi(0)\,$(nm) & $\ell $ (nm) & $\lambda (0)$ (nm)
\\
\hline
{\it full square} & 4.16 & 1.361 & 0.53 & 3.9 & 92 &
7 & 140
\\ \hline
{\it 4-antidot} & 4.16 & 1.353 & 0.53 & 3.9 & 92 & 7
& 140
\\ \hline
{\it 2-antidot} & 4.58 & 1.369 & 0.57 & 4.15 & 92 &
7 & 140
\\ \hline
\end{tabular}
\vspace{0.5cm}
\caption{Values of several characteristic magnitudes for the
three structures. The field corresponding to one flux
quantum, $\mu _0 H_0$, and the effective areas are obtained
by matching of the position of high field peaks to the
theoretical location for a mesoscopic disk. The
characteristic superconducting lengths for the full square
are obtained from the low field behavior of $T_c(\Phi)$
(Eq.~(\ref{lowfield})) and we assume the same values for the
other structures.}
\label{t1}
\end{table}

\end{document}